\documentclass[twocolumn]{aastex63}
\usepackage{multirow}
\usepackage[figuresright]{rotating}
\usepackage{subfigure}
\usepackage{epic,eepic}
\usepackage{graphicx}
\usepackage{longtable}
\usepackage{float}
\usepackage{lineno}
\usepackage{pifont}
\usepackage{gensymb}
\usepackage{multirow}
\usepackage{amsmath}
\usepackage{color}

\shorttitle{}
\shortauthors{Xu et al.}

\begin{document}

\title{Unveiling the nature of G6096: a likely hierarchical triple system}

\correspondingauthor{Xinlin Zhao}
\email{xlzhao@cqu.edu.cn}

\author{Yinghao Xu}
\affiliation{Department of Physics and Chongqing Key Laboratory for Strongly Coupled Physics, Chongqing University, Chongqing, 401331, China}

\author[0009-0005-5459-7433]{Xinlin Zhao}
\affiliation{Department of Physics and Chongqing Key Laboratory for Strongly Coupled Physics, Chongqing University, Chongqing, 401331, China}

\author[0000-0003-3116-5038]{Song Wang}
\affiliation{National Astronomical Observatories, Chinese Academy of Sciences, Beijing 100101, China}
\affiliation{Institute for Frontiers in Astronomy and Astrophysics, Beijing Normal University, Beijing 102206, China}

\author[0000-0001-6753-4611]{Guang-Yao Xiao}
\affiliation{Tsung-Dao Lee Institute, Shanghai Jiao Tong University, Shanghai, 201210, China}
\affiliation{School of Physics and Astronomy, Shanghai Jiao University, Shanghai, 200240, China}

\author[0000-0001-9576-1870]{Zikun Lin}
\affiliation{Department of Astronomy, Xiamen University, Xiamen, Fujian 361005, People’s Republic of China}

\author[0009-0007-4501-4376]{Xue Li}
\affiliation{School of Astronomy and Space Science, University of Chinese Academy of Sciences, Beijing 100049, China}

\author[0000-0002-2912-095X]{Hao-Bin Liu}
\affiliation{Department of Astronomy, Xiamen University, Xiamen, Fujian 361005, People’s Republic of China}

\author[0000-0003-3474-5118]{Henggeng Han}
\affiliation{National Astronomical Observatories, Chinese Academy of Sciences, Beijing 100101, China}

\author[0009-0003-6244-8847]{Weiyi Chen}
\affiliation{School of Astronomy and Space Science, University of Chinese Academy of Sciences, Beijing 100049, China}
\affiliation{National Astronomical Observatories, Chinese Academy of Sciences, Beijing 100101, China}

\author[0009-0004-1088-2139]{Yucong Weng}
\affiliation{School of Astronomy and Space Science, University of Chinese Academy of Sciences, Beijing 100049, China}
\affiliation{National Astronomical Observatories, Chinese Academy of Sciences, Beijing 100101, China}

\author[0000-0001-9037-6180]{Meng Sun}
\affiliation{National Astronomical Observatories, Chinese Academy of Sciences, Beijing 100101, China}
\affiliation{Center for Interdisciplinary Exploration and Research in Astrophysics (CIERA), Northwestern University, Evanston, 60201, USA.}

\author[0000-0002-2419-9590]{Xiaohong Yang}
\affiliation{Department of Physics and Chongqing Key Laboratory for Strongly Coupled Physics, Chongqing University, Chongqing, 401331, China}

\author[0000-0002-2874-2706]{Jifeng Liu}
\affiliation{National Astronomical Observatories, Chinese Academy of Sciences, Beijing 100101, China}
\affiliation{School of Astronomy and Space Science, University of Chinese Academy of Sciences, Beijing 100049, China}
\affiliation{New Cornerstone Science Laboratory, National Astronomical Observatories, Chinese Academy of Sciences, Beijing 100101, China}

\begin{abstract}


G6096 (Gaia DR3 609651611028044544) was recently reported as a wide ($P\sim 450$ days) and eccentric ($e\sim0.18$) binary possibly hosting a massive white dwarf or neutron star.
In this work, through analyses of the projected rotational velocity between the blue and red bands, spectral disentangling, joint radial velocity and astrometric fitting, and X-ray emission, we suggest that the system contains additional visible component(s) rather than a compact object.
We develop a new approach to reveal the nature of G6096 by jointly modeling the spectral energy distribution, rotational velocity, and astrometric measurements. 
Finally, we speculate that G6096 is a hierarchical triple main-sequence star system, comprising a primary with a mass of $\sim 0.75\,M_\odot$ orbited by an inner binary consisting of two dwarfs with masses of $\sim 0.62\,M_\odot$ and $\sim 0.40\,M_\odot$, respectively. This method may help reveal a population of triple systems when applied to {\it Gaia} astrometric data, particularly the upcoming DR4.

\end{abstract}


\section{Introduction}

Searching for compact objects and constructing a complete mass distribution provides a crucial window into stellar evolution and the mechanisms of supernova explosions.
In the era of multi-messenger astronomy, radial velocity (RV) monitoring and astrometry are widely used in the search for compact objects.
Compared to traditional X-ray observations, these two methods can detect compact objects in quiescent binary systems, which typically lack significant X-ray emission from an accretion disk.
Consequently, RV monitoring and astrometry provide a distinct advantage in rapidly growing the sample of compact objects and enabling the construction of a complete mass distribution.
For instance, the number of black hole (BH) binaries confirmed via RV monitoring and astrometry \citep{2014Natur.505..378C,2022NatAs...6.1085S,2022A&A...664A.159M,2023MNRAS.518.1057E,2023MNRAS.521.4323E,2024A&A...686L...2G,2024NatAs...8.1583W} has already reached half the total number discovered through X-ray observations over the past 60 years.
Moreover, RV monitoring and astrometry has also revealed a population of binary systems containing a neutron star (NS) \citep{2022ApJ...940..165Y,2022NatAs...6.1203Y,2024OJAp....7E..58E,2024OJAp....7E..27E}.

The {\it $nss\_two\_body\_orbit$} (NTBO) table in {\it Gaia} Data Release 3 (DR3), derived using the non-single-star (NSS) model \citep{2023A&A...674A...1G}, has provided orbital parameters such as orbital period, eccentricity, and RV semi-amplitude for 443,205 binary systems.
Based on this table, several wide binary systems hosting NSs or BHs have been identified through RV monitoring or astrometry \citep{2023MNRAS.518.1057E,2023MNRAS.521.4323E,2024A&A...686L...2G,2024OJAp....7E..58E,2024OJAp....7E..27E,2024NatAs...8.1583W}.
However, hierarchical triple systems can be a significant source of contamination for compact binaries.
In such systems, the total luminosity of the inner binary may be lower than that of a single star with the same total mass \citep{2019MNRAS.487.5610S,2023MNRAS.518.2991S}, leading them to be misclassified as compact binaries.
%
Some unresolved triple systems have also been identified in recent studies \citep{2023A&A...670A..75C,2024A&A...692A.247B} using NTBO table.
Triple systems are common in the Milky Way, and the fraction of stars residing in such systems increases with their mass \citep{2010ApJS..190....1R,2014AJ....147...86T,2017ApJS..230...15M,2019AJ....157..216W,2021MNRAS.507.3593M,2023ASPC..534..275O}. 
%
%
Based on the Kozai–Lidov mechanism \citep{1962AJ.....67..591K,1962P&SS....9..719L} and stellar evolution, the evolution of triple systems provides a crucial formation channel for different types of binaries \citep{2016ComAC...3....6T,2020A&A...640A..16T,2022A&A...661A..61T,2023ApJ...955L..14S,2024ApJ...970L..11H,2025ApJ...983..115S,2025DDA....5650101S,2025PASP..137g4201S,2026ApJ..1000L..17S,2026enap....2..279P}.

Recently, \cite{2026arXiv260205421L} identified a sample of compact binary candidates using RV data from the Large Sky Area Multi-Object Fiber Spectroscopic Telescope (LAMOST).
Among their sample, four sources were classified as “Class A”, strongly suggesting their companions were likely compact objects.
In these four systems, the wide and eccentric binary {\it Gaia} ID 609651611028044544 (R.A. = 131.0992942$^o$; Decl. = 15.4225867$^o$; hereafter G6096) was proposed to host a massive white dwarf (WD) or an NS with a minimum mass of $0.82 M_{\odot}$.
Under the assumption of a compact binary scenario, G6096 is likely a rare system composed of a pre-main-sequence (PMS) star and a newborn NS.
However, in this work, we demonstrated that the companion in G6096 is actually visible, based on the measurements of projected rotational velocity ($v{\rm sin}i$) and spectral disentangling of the observed spectra.
Finally, combining RV, spectral energy distribution (SED), and astrometric data, we confirmed that G6096 is a hierarchical triple system with an outer K-type MS star and an inner binary consisting of two M dwarfs.
The paper is organized as follows.
In Section \ref{sec:obs_data}, we detail the spectroscopic observations and RV measurements.
In Section \ref{sec:visibel}, we present the properties of the visible star in G6096.
In Section \ref{sec:nature}, we estimate the orbital parameters of G6096 and provide a detailed discussion of its nature.
Finally, we present a summary in Section \ref{sec:summary}.

\section{Observation and data reduction}
\label{sec:obs_data}

LAMOST, a reflecting Schmidt telescope, has provided extensive spectroscopic data that enable the search for compact objects via RV monitoring. 
For instance, more than 40 million optical spectra are released in the LAMOST DR13.
In addition, LAMOST conducts both low- and medium-resolution spectroscopic observations, with spectral resolutions of $\Delta \lambda / \lambda \sim$ 1800 and $\sim$ 7500, respectively. 
The low-resolution spectra (LRS) span a wavelength range from 3690 \AA\ to 9100 \AA\ \citep{2015RAA....15.1095L}, while the medium-resolution spectra (MRS) cover 4950 \AA\ to 5350 \AA\ in the blue arm and 6300 \AA\ to 6800 \AA\ in the red arm \citep{2020arXiv200507210L,2021RAA....21..292W}.

G6096 was observed by LAMOST between April 2015 and April 2021, yielding 8 LRS and 61 MRS. 
We selected spectra with a signal-to-noise ratio (S/N) greater than 5 and calculated the RV for each spectrum using the cross-correlation function (CCF). 
The wavelength range of 4800–5500 \AA\ was used for the LRS, while the red band (excluding the first and last 100 data points) was adopted for the MRS.
The RV value was estimated from the velocity corresponding to the maximum correlation coefficient.
%
%
Table \ref{rvdata.tab} lists the RV measurements of G6096.
Systematic errors in the RV measurements were neglected, as they are expected to be small \citep{2021RAA....21..292W,2024NatAs...8.1583W,2024ApJ...977..245Z}.

\section{Properties of the visible star}
\label{sec:visibel}

\subsection{Distance and extinction}

\begin{figure}
    \center
    \includegraphics[width=0.48\textwidth]{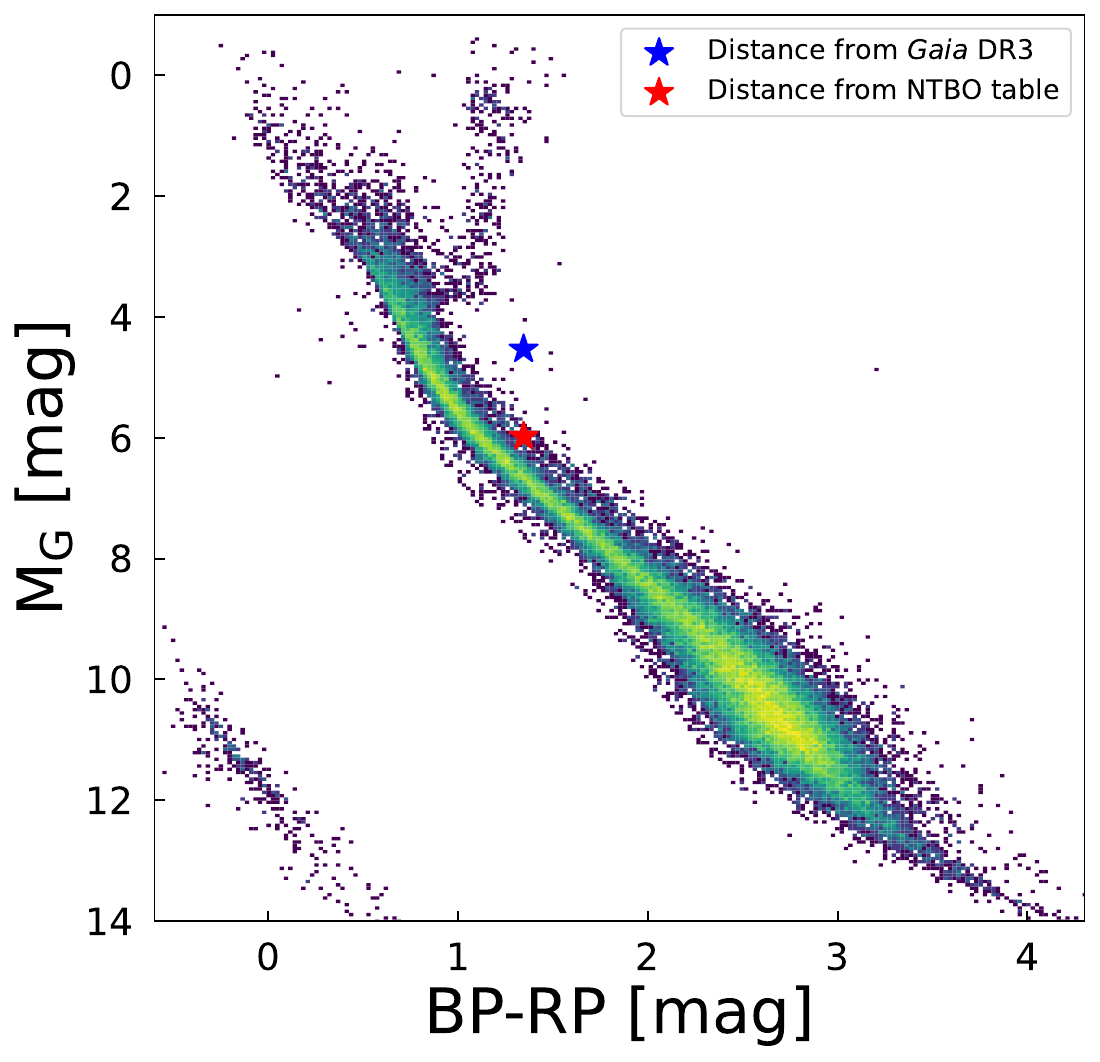}
    \caption{Position of G6096 on the HR diagram. The background stars are plotted for a comparison, which are from {\it Gaia} EDR3 with distances $d <$ 100 pc, $G_{\rm mag}$ between 4--16 mag, and galactic latitudes $|b|$ $>$ 40 deg. WDMS binaries were excluded using the region cuts provided by \cite{2021MNRAS.506.5201R}.} 
    \label{hrd.fig}
\end{figure}

{\it Gaia} DR3 reported a re-normalised unit weight error (RUWE) of 4.144 for G6096 \citep{2022yCat.1355....0G}, suggesting the distance derived from {\it Gaia} DR3 may be unreliable.
Using NSS orbital models, the {\it Gaia} DR3 NTBO table re-derived the orbital parameters of G6096 \citep{2023A&A...674A...1G}, yielding a new parallax of $\varpi=3.2437\pm0.0546$ mas.
We adopted the analytic fitting function from \cite{2025OJAp....8E..62E} to re-derive the parallax uncertainty, yielding a distance of $308\pm18$ pc.
According to this distance, we estimated the reddening $E(B-V)$ using the all-sky three-dimensional dust map \citep{2025ApJS..280...15W}, corresponding to a extinction of $A_V=0.04$ mag.
Figure \ref{hrd.fig} shows the position of G6096 on the Hertzsprung–Russell (HR) diagram, where it appears significantly brighter than MS stars with similar temperature.

\subsection{Atmospheric parameters and SED fitting}

Motivated by previous studies \citep{2023A&A...674A...1G,2026arXiv260205421L}, we investigated the nature of the visible star in G6096 under the assumption of a compact binary.
However, as discussed in Section \ref{sec:compact_binary}, we found that G6096 is unlikely to be either a compact binary or an ordinary binary system.

Assuming the visible star is a single star, we estimated both the spectroscopic and evolutionary masses of the visible star.
First, based on the atmospheric parameters measured by \cite{2019ApJS..245...34X}, we derived final values by weighting multiple observations with the square of S/N, yielding an effective temperature of $T_{\rm eff}=4701\pm23$ K, surface gravity log$g=4.46\pm0.07$, and metallicity [Fe/H]$=-0.40\pm0.08$.
Using the distance from NTBO table, we performed a SED fitting for G6096, obtaining a radius of $R = 1.05^{+0.04}_{-0.04} R_{\odot}$. 
This value is significantly lower than that ($\sim 1.87 R_{\odot}$) reported by \cite{2026arXiv260205421L}, which was derived using the distance from {\it Gaia} DR3.
Combining the updated radius $R = 1.05^{+0.04}_{-0.04} R_{\odot}$ with log$g=4.46\pm0.07$, we derived a spectroscopic mass of $M = 1.14^{+0.22}_{-0.19} M_{\odot}$ for the visible star.
Second, we estimated the evolutionary mass and radius of the visible star using the {\it isochrones} Python package \citep{2015ascl.soft03010M}. 
Based on the observed atmospheric parameters and photometric data, the {\it isochrones} analysis yielded an evolutionary mass and radius of $M = 0.88^{+0.02}_{-0.02} M_{\odot}$ and $R = 1.01^{+0.02}_{-0.02}\,R_{\odot}$, consistent with the spectroscopic mass and the radius from SED fitting.  
%

\section{Nature of the system}
\label{sec:nature}

\subsection{RV Fitting}

\begin{figure*}[ht!]
    \centering
    \includegraphics[width=0.98\textwidth]{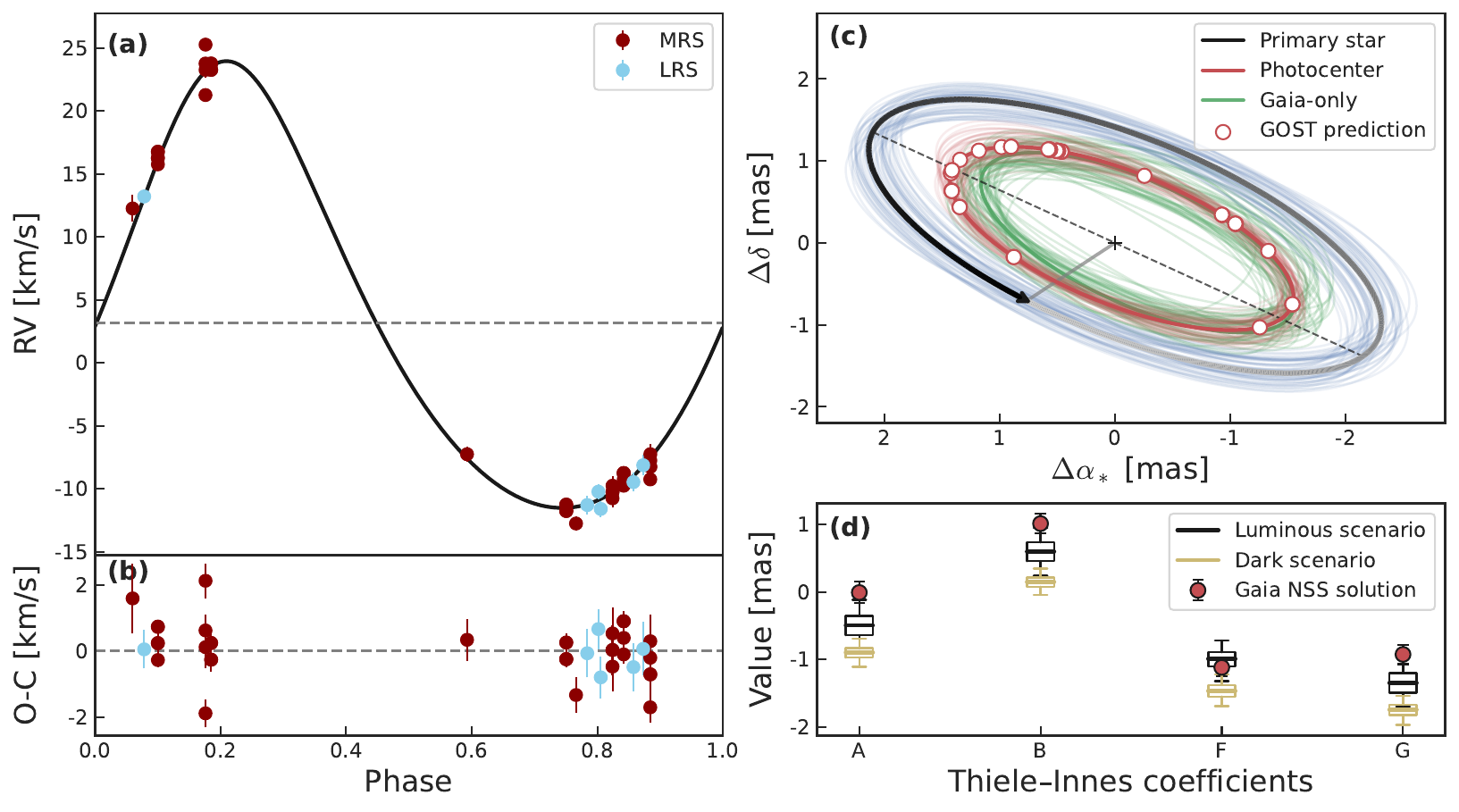}
    \caption{Results of {\it The Joker} and the joint fitting of RV and astrometry. Panel (a): Keplerian orbital fit from {\it The Joker}. Panel (b): RV residuals. Panel (c): astrometric orbits of the primary star (black), photocenter (red) and {\it Gaia}-only solution (green) in the sky-projected plane. The black dashed line inside the orbit is the line of nodes joining the ascending node and the descending node. The plus symbol denotes the system's barycenter, and the grey line connects it with the periapsis. The arrow indicates the direction of the star's orbital motion. {\it Gaia} observational times simulated with GOST are indicated by hollow dots. Panel (d): A comparison between the TI coefficients obtained from {\it Gaia} measurements and the predictions of two scenarios. The inner thick line, the body and the edge of the boxplot respectively indicate the median, $1\,\sigma$ uncertainty and $3\,\sigma$ uncertainty. 
\label{fig:RV_Gaia_res}}
\end{figure*}

We performed a Keplerian orbital fit using the re-calculated RV measurements listed in Table \ref{rvdata.tab}.  
Each measurement was treated as an independent observation.
The fit was carried out with {\it The Joker} \citep{2017ApJ...837...20P}, a Python package designed to infer Keplerian orbits of binary systems via a custom Markov Chain Monte Carlo (MCMC) sampler.  
The resulting orbital solution yielded the following parameters:  
orbital period $P = 447.18^{+3.35}_{-3.16}$ d,  
eccentricity $e = 0.18^{+0.03}_{-0.02}$,  
argument of periastron $\omega = 5.98^{+0.16}_{-0.24}$,  
mean anomaly at the epoch of the first exposure $M_0 = 0.73^{+0.13}_{-0.14}$,  
RV semi-amplitude $K = 17.74^{+0.16}_{-0.14}$ km/s,  
and systemic RV $\nu_0 = 3.17^{+0.57}_{-0.62}$ km/s.  
This orbital solution is consistent with the results from the NTBO table and \cite{2026arXiv260205421L}, confirming the reliability of {\it The Joker} analysis.  
Figure \ref{fig:RV_Gaia_res} shows the folded RV data and the RV curve given by {\it The Joker}.

Consequentially, we calculated the mass function $f(M_2)$ using these parameters as follows,
\begin{equation}
    f(M_{2}) = \frac{M_{2} \, \textrm{sin}^3 i} {(1+q)^{2}} = \frac{P \, K_{1}^{3} \, (1-e^2)^{3/2}}{2\pi G},
\label{eq5}
\end{equation}
\noindent
where $M_{1}$ and $M_{2}$ are the masses of the visible star and companion. $q = M_{1}/M_{2}$ is the mass ratio, and $i$ is the inclination.
Finally, the mass function of G6096 is $f(M_2) = 0.246^{+0.008}_{-0.008} M_{\odot}$.

\subsection{An unlikely scenario: a PMS + NS binary}
\label{sec:compact_binary}

\begin{figure}
    \center
    \includegraphics[width=0.48\textwidth]{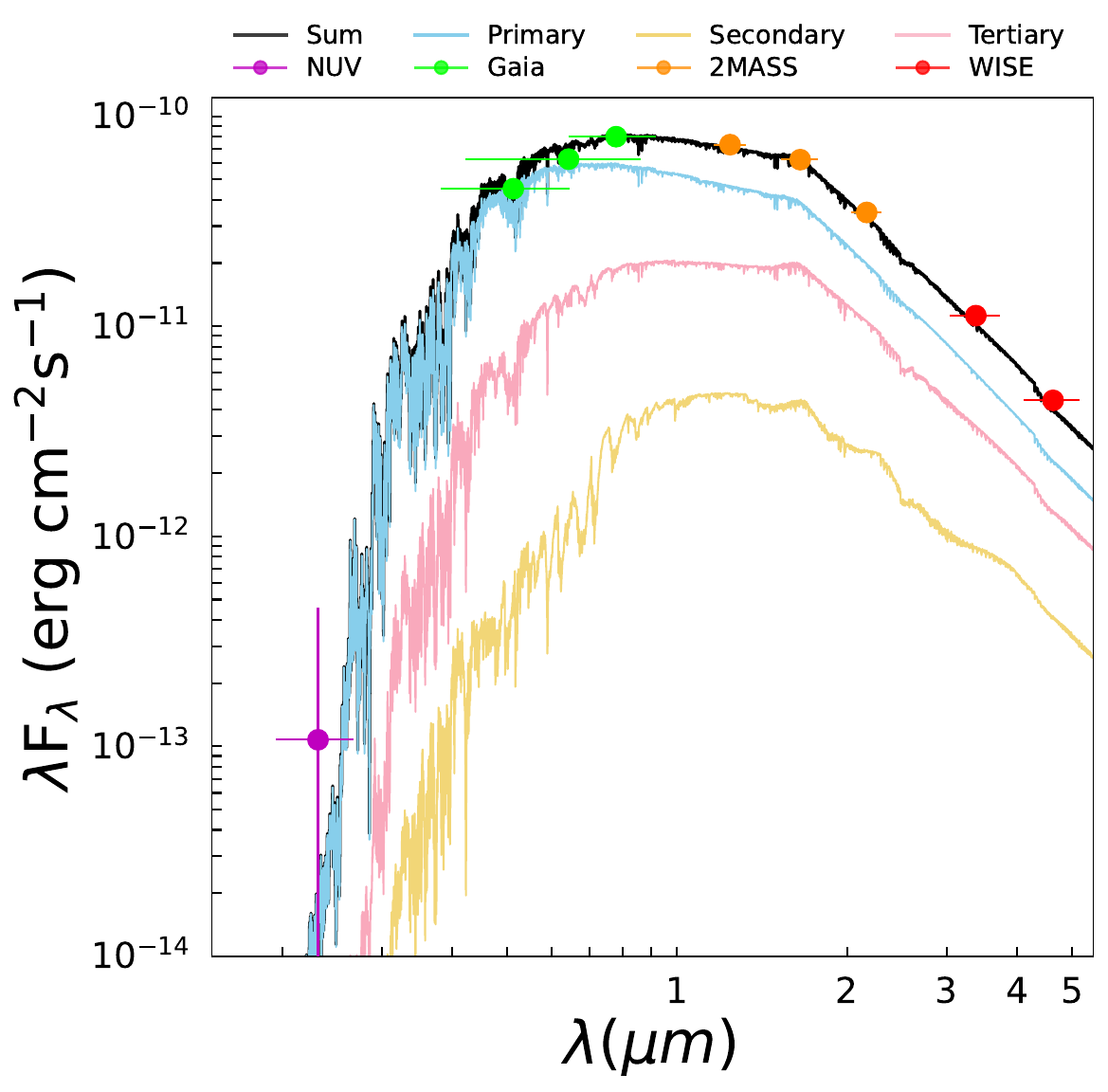}
    \caption{SED fitting of G6096 from joint SED and dynamical results. The black line shows the best-fitting. The blue, yellow, and pink lines represent the theoretical spectra of the primary, secondary, and tertiary, respectively. The purple, green, orange, and red dots denote the observed magnitudes in the NUV, {\it Gaia}, {\it 2MASS}, and {\it WISE} bands, respectively.} 
    \label{sed_fitting.fig}
\end{figure}

Combining the mass function ($f(M_2) = 0.246^{+0.008}_{-0.008} M_{\odot}$) and the inclination ($64.41^{\circ}\pm2.24^{\circ}$) from NTBO table \citep{2023A&A...674A...1G}, we obtained a companion with the masses of $1.24^{+0.12}_{-0.12} M_{\odot}$ and $1.09^{+0.04}_{-0.04} M_{\odot}$ for the spectroscopic ($1.14^{+0.22}_{-0.19} M_{\odot}$) and evolutionary ($0.88^{+0.02}_{-0.02} M_{\odot}$) mass of the visible star, respectively.
Consequently, if the companion is indeed a compact object, it would be either a massive WD or an NS.
However, no clear near-ultraviolet (NUV) emission is found (Figure \ref{sed_fitting.fig}), reducing the possibility of a young/hot WD companion.
Moreover, the evolutionary stage inferred from {\it isochrones} indicated that the visible star in G6096 is a PMS star with an estimated age of approximately $15.36^{+1.87}_{-1.87}$ Myr.
Finally, under the assumption that G6096 is a compact binary, this would make it a system comprising a visible PMS and a likely newborn NS.
However, several aspects suggest that this scenario is unlikely, as shown in the following analysis.

\subsubsection{Projected rotational velocity}
\label{sec:vsini}

For compact binaries, spectral line broadening should be the same across all wavelength ranges, as the spectra is entirely contributed by the visible star. 
Consequently, discrepancies in spectral line profiles between the red and blue bands provide insight into the intrinsic nature of binaries.

To assess whether an additional visible component may be present in the observed spectra of G6096, we measured $v{\rm sin}i$ separately in the blue (5000--5300 \AA) and red (6300--6700 \AA) bands using MRS spectra.
%
We first adopted a PHOENIX\footnote{\url{https://phoenix.astro.physik.uni-goettingen.de}} model with atmospheric parameters ($T_{\rm eff}=4700$ K, log$g=4.5$, and [Fe/H]$=-0.5$) similar to those of G6096 as the template.
A series of rotationally broadened templates were generated by convolving this model with rotational kernels spanning $v{\rm sin}i$ from 0 to 120 km/s in steps of 1 km/s.
The best-fit $v{\rm sin}i$ value was then determined by minimizing the $\chi^{2}$ between the observed spectrum and the broadened templates.
The reliability of the results was validated in Figure \ref{vsini_test.fig}.
For the blue band, we obtained $v{\rm sin}i = 45.5\pm14.4$ km/s, whereas the red band yielded a $v{\rm sin}i$ value consistent with zero.
Figure \ref{best_fit_vsini.fig} displays the best-fit result.
As in the case of Unicorn \citep{2022MNRAS.512.5620E}, the discrepancy may arise from a cooler companion whose contribution varies between the blue and red bands.
%
Moreover, most K-type MS stars have a $v{\rm sin}i$ value of $\sim 10$ km/s \citep{2010A&A...520A..15M}, significantly lower than the value derived from the blue band. 
This inconsistency could also be attributed to the presence of a cooler companion (Section \ref{sec:sed_fap_joint_fit}).


\subsubsection{Spectral disentangling}
\label{sec:sd}

We applied the spectral disentangling algorithm proposed by \cite{1994A&A...281..286S} to search for a possible faint companion in the LAMOST MRS spectra.
Under the assumption that the component spectra remain constant over time, the algorithm derives two effective spectral components (e.g., Component A and Component B).

We first generated a grid of synthetic binary spectra with mass ratios of $q=0.8, 0.9, 1.0$ and $v{\rm sin}i$ of $50, 100, 150, 200$ km/s to assess the detection limits of the spectral disentangling algorithm for G6096.
The synthetic spectra are constructed following \citet{2022MNRAS.517..356K}:
\begin{equation}
    {f}_{\lambda,{\rm binary}}=\frac{{f}_{\lambda,2} + k_\lambda {f}_{\lambda,1}}{1+k_\lambda},
	\label{eq6}
\end{equation}
where $k_\lambda= \frac{B_\lambda(T{_{\rm eff,1}})~M_1}{B_\lambda(T{_{\rm eff,2}})~M_2} 10^{{\rm log}g_2-{\rm log}g_1}$ is the flux ratio and $B_\lambda$ is the Plank function.
Spectral disentangling was performed on these synthetic binary spectra over the wavelength range of 5000--5150 \AA.
Visual inspection revealed that all companions can be successfully recovered, suggesting the feasibility of this spectral disentangling algorithm for G6096.

\begin{figure*}
    \center
    \includegraphics[width=0.98\textwidth]{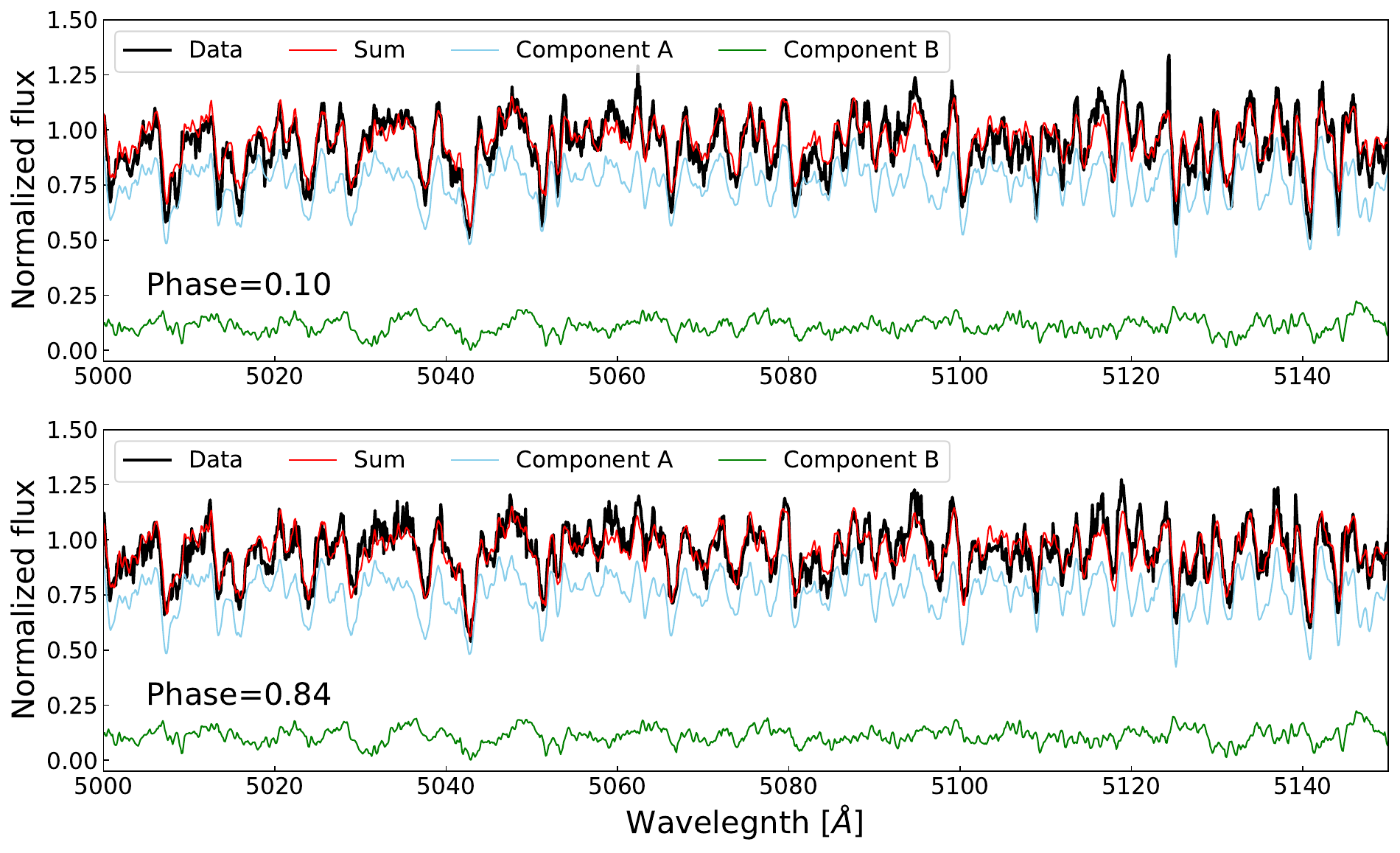}
    \caption{Spectral disentangling results for $q =$ 0.92 (=1.14$/$1.24) at phases of large redshift and blueshift. The blue lines mark the reconstructed spectra of the visible star, while the green lines represent the second component in each spectra. The red lines are the sum of the two components, and the black lines represent the observed spectra.} 
    \label{dis_6096.fig}
\end{figure*}

Assuming a normal binary system, we derived a $G$-band flux ratio of $\approx$7 based on the mass ratio of $q = 0.92$ (=1.14$/$1.24). 
Using this flux ratio, we applied the spectral disentangling method to the LAMOST MRS spectra of G6096 in the wavelength range of 5000--5150 \AA. 
Figure \ref{dis_6096.fig} presents the disentangling results. 
Several absorption features can be seen in Component B (green lines in Figure \ref{dis_6096.fig}), reducing the probability that G6096 hosts a compact companion.

\subsubsection{Inconsistency between RV and astrometric data}
\label{sec:rv_astrometric_data}

The astrometric orbit of G6096 was derived from the NTBO table utilizing the observed astrometric data. 
Under the dark companion assumption, this astrometric orbit predicts an RV semi-amplitude of $11.10\pm0.42$ km/s, significantly inconsistent with the $K$ value derived from the RV fitting.
Since the RV fitting and NTBO table yielded nearly consistent values for the orbital period $P$ and eccentricity $e$, this discrepancy might be attributed to the neglect of light contamination from the companion. 
Thus, we conducted a joint analysis of the RV and {\it Gaia} astrometry (Section \ref{sec:rv_astromotry}) to investigate whether the companion of G6096 could be an unseen object.
As a result, the astrometric data yielded a poor fit under the dark companion assumption (Figure \ref{fig:RV_Gaia_res}), disfavoring the presence of an NS.

\subsubsection{X-ray observation}  
\label{sec:Xray}

The G6096 was covered by the XMM-Newton Slew Survey \citep{Jansen2001,Saxton2008} in nine observation segments, but without a detection of G6096.
We downloaded the corresponding 0.2--12 keV Pipeline Processing System (PPS) images and exposure maps from the XMM-Newton Science Archive (XSA). 
Combining all available slew images and exposure maps, we estimated the upper limit using the {\it eupper}\footnote{\url{https://xmm-tools.cosmos.esa.int/external/sas/current/doc/eupper/index.html}} task in the XMM-Newton Science Analysis System \citep[SAS;][]{Gabriel2004}. 
A circular source aperture of 30$^{\prime\prime}$ was adopted at the target position, and the background was estimated from an annulus with inner and outer radii of 60$^{\prime\prime}$ and 180$^{\prime\prime}$, respectively.
We adopted an encircled energy fraction (EEF) of 0.85, appropriate for EPIC-pn slew data with a 30$^{\prime\prime}$ aperture \citep{Ruiz2022}. 
No counts were detected in the source region, while the scaled background contribution was 1.564 counts. 
For a total effective exposure of 68.62 s, we derived a 99\% confidence upper limit of 0.07897 counts s$^{-1}$  \citep{1991ApJ...374..344K}.
The hydrogen column density, $N_{\rm H}$, was estimated using the linear relation between $N_{\rm H}$ and reddening: $N_{\rm H}=M\times10^{21}$ cm$^{-2}$ $E{\rm (B-V)}$.
In this work, we adopted the value $M=5.55$ derived by \citet{1995A&A...293..889P} based on ROSAT observations of the interstellar medium.
Assuming a power-law spectrum with different photon indices (e.g., $\tau=$1.5, 2), we calculated unabsorbed fluxes of 2.543$\times 10^{-13}\ \rm erg\, s^{-1}\, cm^{-2}$ and 1.554$\times 10^{-13}\ \rm erg\, s^{-1}\, cm^{-2}$, respectively, using WebPIMMS\footnote{\url{https://heasarc.gsfc.nasa.gov/cgi-bin/Tools/w3pimms/w3pimms.pl}}. 
Taking the distance of 308 pc, we derived upper limit X-ray luminosities of 2.89$\times$10$^{30}$ $\rm erg\, s^{-1}$ ($\tau=1.5$) and 1.78$\times$10$^{30}$ $\rm erg\, s^{-1}$ ($\tau=2$)). 
Consequently, G6096 likely exhibits no or very weak X-ray emission, ruling out the presence of a young NS \citep{2020MNRAS.496.5052P,2024PhRvD.109l3018D}.

\subsection{A more plausible explanation: a hierarchical triple system}

Based on the results presented in Section \ref{sec:compact_binary}, G6096 is unlikely to be either a compact binary (e.g., PMS + NS) or an ordinary binary system.
Because a normal star with a mass of $\approx$1.24 $M_{\odot}$ is brighter than the visible star, here we investigated the possibility that G6096 is a hierarchical triple system.

\subsubsection{Joint fitting of RV and astrometric analysis} 
\label{sec:rv_astromotry}

We conducted a joint fitting of the RV and {\it Gaia} astrometry to derive a self-consistent Keplerian orbit and uncover the nature of the companion under both the dark companion and luminous companion scenarios.
The dark companion scenario indicates an unseen companion while the luminous companion scenario suggests a visible companion.
Our joint fitting is largely following the methodology developed by \cite{Winn2022AJ} (also see \citealt{El_Badry2023MNRAS}). 
The primary fitted parameters include the orbital period $P$, RV semi-amplitude $K$, eccentricity $e$, argument of periastron $\omega$ of stellar reflex motion, orbital inclination $i$, longitude of ascending node $\Omega$, mean anomaly $M_{0}$ at J2016.0, and parallax $\varpi$. 
Uniform priors are adopted for these parameters. 
With total likelihood $\mathcal{L}=\mathcal{L}_{\rm RV}\cdot\mathcal{L}_{\rm gaia}$, the orbital solution (Section \ref{sec:RV_Astrometry}) was derived by sampling the posterior using the parallel-tempering MCMC sampler \texttt{ptemcee} \citep{Vousden2016}. 
The setup employed 5 temperatures, 100 walkers, and 50,000 steps per chain to generate posterior distributions for all fitting parameters, with the first 20,000 steps discarded as burn-in.

Table \ref{Tab:result} lists best-fit results for both the dark and luminous companion scenarios. 
We found that the best-fit solution in the dark companion scenario did not match the observed Thiele–Innes (TI) coefficients (e.g., $A$, $B$, $F$, $G$) well. 
In contrast, the best-fit results under the luminous companion scenario were consistent with the observed TI coefficients within 3 $\sigma$.
To quantify the difference between the two best-fit solutions, we calculated the Bayesian Information Criterion difference ($\rm \Delta BIC$) \citep{Schwarz1978} between the dark and luminous companion scenarios. 
Ultimately, we obtained a $\rm \Delta BIC$ of approximately 30, indicating the presence of a luminous companion provides a better explanation for the observed RV and astrometric data. 
Figure \ref{fig:RV_Gaia_res} displays the best-fit results under both the dark and luminous companion scenarios.

\subsubsection{Joint fitting of SED and dynamical results}
\label{sec:sed_fap_joint_fit}

In the luminous companion scenario, the astrometric model yields a function $f_{\rm ap}$ defined as follows:
\begin{equation}
\label{equ:fap}
f_{\rm ap} = (1-\frac{m_\star\,f_c}{m_c\,f_\star})(1+\frac{f_c}{f_\star})^{-1},  
\end{equation}
where $m_\star$ and $m_c$ are the masses of primary and companion, and $f_\star$ and $f_c$ are their fluxes in $G$-band.
For G6096, the joint RV and astrometric analysis derived a value of $f_{\rm ap}=0.668^{+0.072}_{-0.055}$ (Table \ref{Tab:result}).
This $f_{\rm ap}$ can be used to infer the flux ratio for different mass sets of primary and companion. 
Assuming a primary with a mass ranging from 0.2 to 1 $M_{\odot}$, the corresponding mass of the companion ranges from 0.6 to 1.2 $M_{\odot}$.
The flux ratio derived from the $f_{\rm ap}$ value is approximately 0.22 to 0.33.
However, under the assumption of a normal binary system, the flux ratio predicted by stellar evolution models would be 2 to 34.
This discrepancy suggests that an ordinary binary configuration cannot adequately explain the observed properties of G6096.
Alternatively, a multiple-star system, especially a hierarchical triple, may provide a plausible explanation for the nature of G6096.

In the astrometric model, $f_{\rm ap}$ is derived primarily from orbital parameters and is independent of the mass of the primary.
To investigate the possibility of a triple system, we performed a joint fit to the observed SED of G6096 and the $f_{\rm ap}$ value obtained from the joint RV and astrometric analysis.
In this framework, the primary was treated as the outer star with mass $M_1$ and radius $R_1$.
The inner binary consists of two stars with masses $M_2$, $M_3$ and radii $R_2$, $R_3$, respectively.
Based on stellar evolutionary tracks, we generated a set of hierarchical triple systems spanning a range of mass combinations for $M_1$, $M_2$, and $M_3$.
These models were employed to perform the joint fitting (Appendix \ref{sec:joint_fit_sed_fap}).
Table \ref{joint_fitting.tab} lists the best-fit results.
Consequently, we obtained $M_1 = 0.75^{+0.01}_{-0.01} \, M_\odot$ and $R_1 = 0.73^{+0.01}_{-0.01} \, R_\odot$ for the primary star, and $M_2 = 0.62^{+0.01}_{-0.01} \, M_\odot$, $M_3 = 0.40^{+0.01}_{-0.01} \, M_\odot$, $R_2 = 0.62^{+0.01}_{-0.01} \, R_\odot$, and $R_3 = 0.40^{+0.03}_{-0.03} \, R_\odot$ for the inner binary components.
The best-fit model yielded an estimated $f_{\rm ap}=0.668^{+0.018}_{-0.021}$, consistent with the value derived from the joint fitting of RV and astrometry.
Figure \ref{sed_fitting.fig} shows the best-fitting for the observed SED.
Compared with the compact binary or ordinary binary scenarios, we concluded that G6096 is most likely a hierarchical triple system, which can account for all the observed data.

The joint fitting returns a near-solar metallicity, significantly deviating from the value estimated using spectral templates \citep{2019ApJS..245...34X}.
This discrepancy is expected. 
The observed spectrum of G6096 is a composite of contributions from the primary, secondary, and tertiary, leading to inaccurate abundance estimates when using single-star spectral templates.
Based on the atmospheric parameters of the primary, secondary, and tertiary derived from the joint fitting, we synthesized a theoretical spectrum to estimate the $v{\rm sin}i$ value.
Using this spectrum, we obtained a $v{\rm sin}i$ of $\sim 20$ km/s in the blue band, significantly higher than the typical value of K-type MS stars.
Two other reasons can also contribute to the measured $v{\rm sin}i$ value in the blue band:
(1) the orbital motion of the inner binary may broaden the spectral line profiles, leading to a larger measured $v{\rm sin}i$; (2) The lower S/N in the blue band results in larger uncertainty in the $v{\rm sin}i$ measurement.

\begin{table}
\caption{The best-fit results from the joint fitting of SED and dynamical results. The primary is the outer star, while the secondary and tertiary are the hotter and cooler components of the inner binary, respectively. \label{joint_fitting.tab}}
\centering
\renewcommand{\arraystretch}{1.5}
\setlength{\tabcolsep}{1.mm}
 \begin{tabular}{lcccc}
\hline\noalign{\smallskip}
Parameters & System & Primary & Secondary & Tertiary  \\
\hline\noalign{\smallskip}
Age [Gyr] & $9.97^{+1.67}_{-1.41}$ & & & \\
$f_{\rm ap}$ & $0.668^{+0.018}_{-0.021}$ & & & \\
\textnormal{[Fe/H]} & $0.03^{+0.03}_{-0.04}$ & & & \\
$T_{\rm eff}$ [K] & & $4854^{+11}_{-10}$ & $4028^{+43}_{-39}$ & $3318^{+21}_{-20}$ \\
$M$ [$M_{\odot}$] & & $0.75^{+0.01}_{-0.01}$ & $0.62^{+0.01}_{-0.01}$ & $0.40^{+0.01}_{-0.01}$ \\
$R$ [$R_{\odot}$] & & $0.73^{+0.01}_{-0.01}$ & $0.62^{+0.01}_{-0.01}$ & $0.40^{+0.01}_{-0.01}$ \\
\noalign{\smallskip}\hline
\end{tabular}
\end{table}

\subsection{Comparison with other triple systems}
\label{sec:discussion}

A population of triple systems has been identified by previous studies using various methods \citep{2010ApJS..190....1R,2012AJ....143..137G,2013ApJ...768...33R,2014AJ....147...86T,2014AJ....147...87T,2024A&A...692A.247B,2025PASP..137i4201S,2026AJ....171...29K}.
Among these confirmed systems, a significant number of triple systems are identified via eclipse timing variations (ETVs) in their light curves \citep{2012AJ....143..137G,2013ApJ...768...33R,2026AJ....171...29K}. 
However, no clear ETV signals can be found in the TESS light curve of G6096, making it difficult to verify the nature of this system and  constrain the orbital period of the inner binary.
Recently, \cite{2026PASJ..tmp...58T} reported a similar triple system, {\it Gaia} DR3 1010268155897156864, using high-resolution optical spectra by analyzing the CCF profiles between residual spectra and template.
However, this approach requires a number of high-resolution observations to detect the spectral features of the inner binary in the residuals. 
Instead, by combining the observed SED, RV, and astrometric data, we developed a new approach to reveal the nature of G6096.

We compared G6096 with the multiple star catalog provided by \cite{2023A&A...670A..75C}, which was constructed by cross-matching the NTBO table with several databases of eclipsing binaries. 
Figure \ref{compare_3b_systems.fig} illustrates the comparison between the inner and outer orbital periods of triple system candidates from previous study \citep{2023A&A...670A..75C}.
All systems reside above the threshold of $P_{\rm out}/P_{\rm in} = 5$, confirming their dynamical stability \citep{2001MNRAS.321..398M}, and as is consistent with wider triples \citep{2025PASP..137i4201S}.
Furthermore, all systems with a “Orbital” solution exhibit a period ratio ($P_{\rm out}/P_{\rm in}$) above 25. 
Consequently, based on the estimated outer orbital period of $ \sim$450 days for G6096, the inner binary is likely a close binary with a period shorter than 18 days.

\section{Summary}  
\label{sec:summary}

\cite{2026arXiv260205421L} reported G6096 as a compact binary candidate possibly harboring either a massive WD or an NS.
Using the RV data from LAMOST, we derived an orbital period of approximately 447 days and an eccentricity of 0.18, consistent with previous findings \citep{2023A&A...674A...1G,2026arXiv260205421L}.
Assuming a compact binary configuration, G6096 could be a rare binary consisting of a PMS and a newborn NS.
However, the significant discrepancy between the $v{\rm sin}i$ values measured in the blue and red bands suggests that G6096 is unlikely to be a compact binary.
This conclusion is further supported by spectral disentangling, joint RV and astrometric fitting, and X-ray analysis.

To investigate the possibility of a triple system for G6096, we performed a joint fit to the observed SED and dynamical results.
The best-fit solution corresponds to a hierarchical triple system in which the primary is a $0.75^{+0.01}_{-0.01} M_{\odot}$ K-type MS star, orbited by an inner pair of M-type dwarfs with masses of $0.62^{+0.01}_{-0.01} M_{\odot}$ and $0.40^{+0.01}_{-0.01} M_{\odot}$.
Compared to the PMS + NS scenario, the triple system is more plausible and is consistent with all current observational constraints.
Based on known triple systems, we inferred that the inner binary of G6096 is likely a close binary with an orbital period shorter than 18 days.

Currently, some triple systems have been discovered through ETVs in their light curves. However, triple systems such as G6096 are difficult to identify via ETVs alone. 
In this work, we proposed a novel approach that combines SED, RV, and astrometric data to uncover such systems. 
By applying this method to {\it Gaia} astrometric data, particularly the upcoming DR4, we expect to reveal a population of potentially valuable triple systems.

\begin{acknowledgements}

This work was funded by the Strategic Priority Program of the Chinese Academy of Sciences under grant number XDB1160302 (Song Wang), the National Key Research and Development Program of China under grant number 2023YFA1607901 (Song Wang), the National Natural Science Foundation of China (NSFC) under grant number 12588202 (Jifeng Liu), the NSFC under grant number 12273057 (Song Wang), science research grants from the China Manned Space Project (Song Wang), the New Cornerstone Science Foundation through the New Cornerstone Investigator Program and the XPLORER PRIZE (Jifeng Liu), Chongqing Natural Science Foundation under grant number CSTB2023NSCQ-MSX0093 (Xiaohong Yang), the NSFC under grand number 12547101 (Xinlin Zhao), and the Postdoctoral Fellowship Program of China Postdoctoral Science (CPSF) under grant number GZB20250739 (Xinlin Zhao).

\end{acknowledgements}

\bibliography{main}{}
\bibliographystyle{aasjournal}

\clearpage
\appendix
\renewcommand*\thetable{\Alph{section}.\arabic{table}}
\renewcommand*\thefigure{\Alph{section}\arabic{figure}}

\section{RV mesurements}  
\label{sec:zprv}

\setcounter{table}{0}

Here we present the RV measurements of G6096 in Table \ref{rvdata.tab}.

 \begin{table}
 \caption{Barycentric-corrected RV values of G6096. \label{rvdata.tab}}
 \setlength{\tabcolsep}{5pt}
\begin{tabular}{ccccc|ccccc}
 \hline\noalign{\smallskip}
 BMJD & RV & Uncertainty & S/N$_{r}$ & Source & BMJD & RV & Uncertainty & S/N$_{r}$ & Source \\
 (day) & (km/s) & (km/s) &  &  & (day) & (km/s) & (km/s) & & \\
 \hline\noalign{\smallskip}
58451.84311 & 12.26 & 1.06 & 5.05 & MRS & 59240.73222 & -10.76 & 0.74 & 6.60 & MRS \\
58469.80146 & 16.26 & 0.19 & 42.84 & MRS & 59240.74889 & -9.76 & 0.78 & 6.68 & MRS \\
58469.81813 & 16.76 & 0.19 & 43.00 & MRS & 59240.76555 & -10.26 & 0.94 & 6.55 & MRS \\
58469.83410 & 16.26 & 0.30 & 43.43 & MRS & 59248.66138 & -8.76 & 0.26 & 31.10 & MRS \\
58469.85077 & 15.76 & 0.21 & 38.80 & MRS & 59248.67735 & -8.76 & 0.30 & 30.01 & MRS \\
58469.86674 & 16.76 & 0.22 & 39.32 & MRS & 59248.69332 & -9.76 & 0.27 & 31.07 & MRS \\
58503.72588 & 21.26 & 0.43 & 12.18 & MRS & 59248.70999 & -9.26 & 0.36 & 29.63 & MRS \\
58503.74186 & 23.76 & 0.48 & 10.45 & MRS & 59248.72596 & -8.76 & 0.32 & 27.83 & MRS \\
58503.75852 & 25.27 & 0.53 & 9.63 & MRS & 59267.61785 & -9.26 & 0.46 & 17.17 & MRS \\
58503.77519 & 23.26 & 0.64 & 7.48 & MRS & 59267.63382 & -8.26 & 0.50 & 13.88 & MRS \\
58507.71691 & 23.76 & 0.25 & 24.11 & MRS & 59267.64979 & -7.75 & 0.46 & 13.36 & MRS \\
58507.73288 & 23.76 & 0.28 & 21.40 & MRS & 59267.66646 & -7.25 & 0.80 & 7.16 & MRS \\
58507.74886 & 23.26 & 0.34 & 19.29 & MRS & 59267.68243 & -8.26 & 0.68 & 10.20 & MRS \\
58507.76552 & 23.26 & 0.36 & 19.90 & MRS & 60031.51848 & -7.25 & 0.63 & 12.37 & MRS \\
59207.77348 & -11.76 & 0.21 & 39.00 & MRS & 57118.51839 & 13.21 & 0.58 & 71.61 & LRS \\
59207.78946 & -11.26 & 0.27 & 31.48 & MRS & 57443.65289 & -11.60 & 0.63 & 60.18 & LRS \\
59207.80613 & -11.76 & 0.24 & 41.62 & MRS & 59222.73930 & -11.29 & 0.73 & 126.53 & LRS \\
59207.82210 & -11.26 & 0.24 & 38.34 & MRS & 59230.67512 & -10.23 & 0.61 & 218.10 & LRS \\
59207.83807 & -11.76 & 0.25 & 34.81 & MRS & 59255.62770 & -9.47 & 0.73 & 157.89 & LRS \\
59214.76136 & -12.76 & 0.55 & 5.89 & MRS & 59262.61258 & -8.13 & 0.81 & 113.43 & LRS \\
59240.71625 & -10.26 & 0.75 & 6.01 & MRS &  &  & &  & \\
 \noalign{\smallskip}\hline
 \end{tabular}
 \end{table}

\section{Validation of the $v{\rm sin}i$ measurement}  
\label{sec:test_for_vsini}

\setcounter{figure}{0}
\setcounter{table}{0}

\begin{figure}
    \center
    \includegraphics[width=0.49\textwidth]{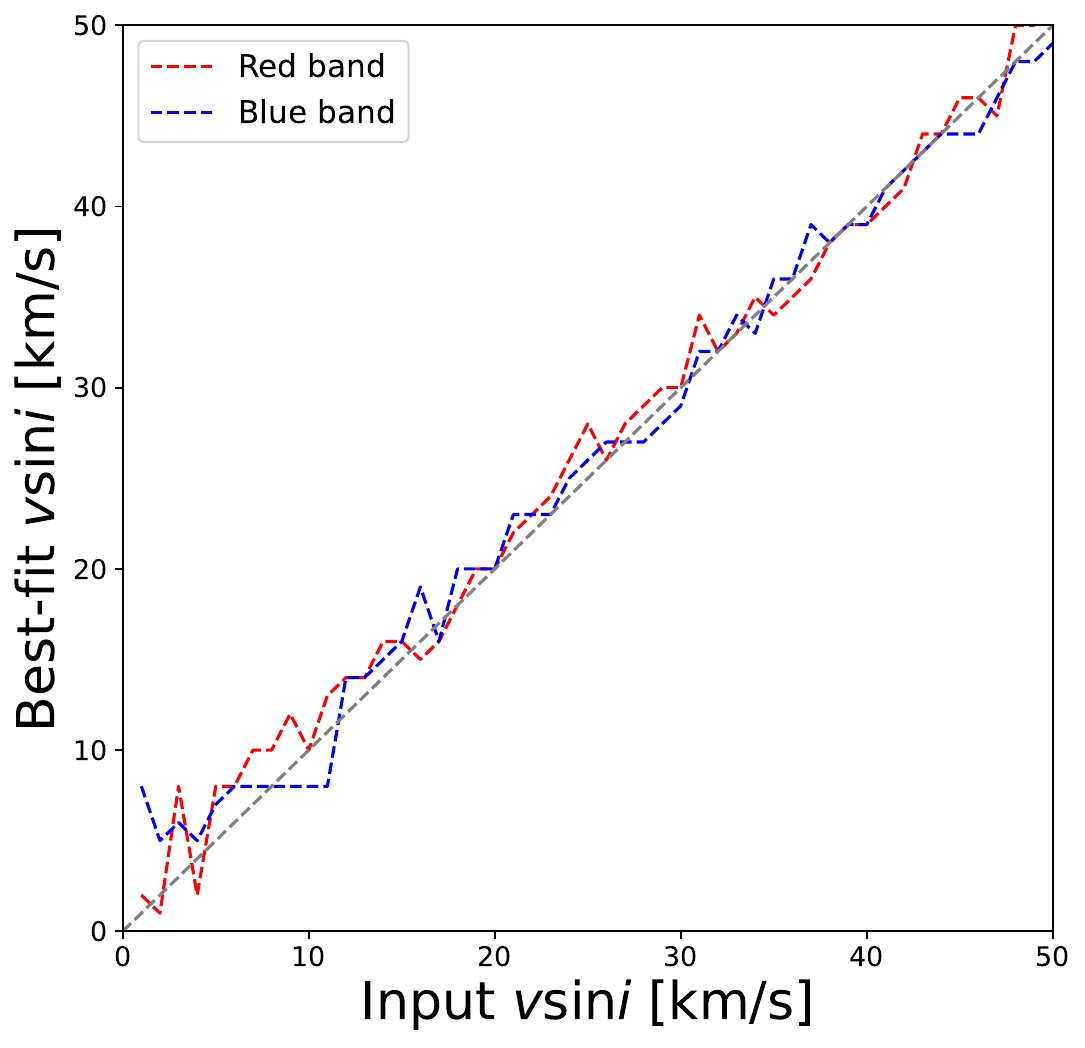}
    \caption{Test results for the $v{\rm sin}i$ measurement. The red dashed line shows the $v{\rm sin}i$ values measured from the red band of the synthetic spectra, and the blue dashed line shows those from the blue band. The gray dashed line represents the 1:1 line.}  
    \label{vsini_test.fig}
\end{figure}

\begin{figure}
    \center
    \includegraphics[width=0.98\textwidth]{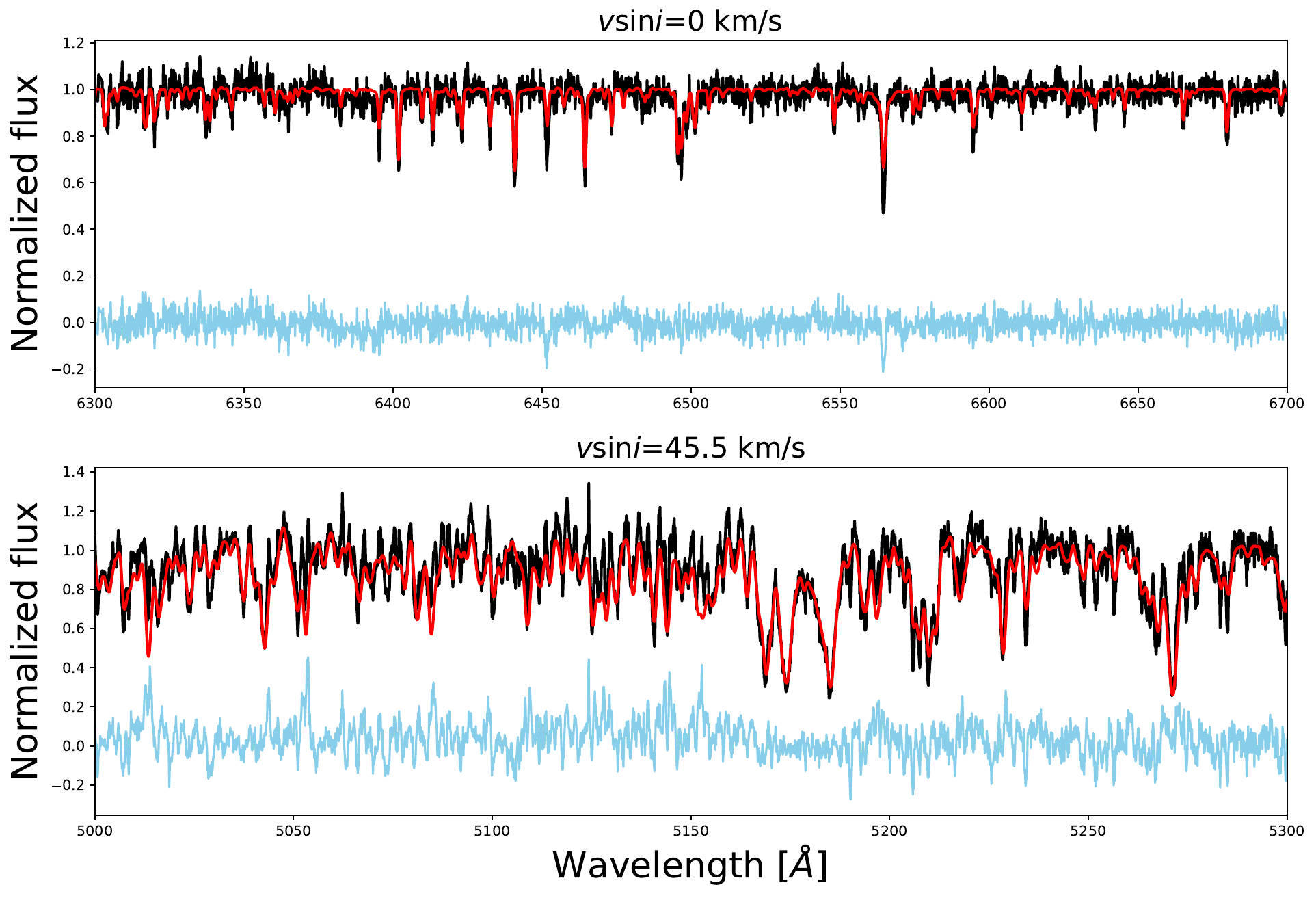}
    \caption{Comparison of the observed spectrum of G6096 with the best-fit template. The black line shows the observed MRS spectrum at maximum redshift, the red line represents the best-fit synthetic template, and the blue line indicates the residual spectrum.}  
    \label{best_fit_vsini.fig}
\end{figure}

The $v{\rm sin}i$ measurement derived from the blue band is comparable to the instrumental velocity of LAMOST MRS.  
To assess the reliability of the $v{\rm sin}i$ measurement for G6096, we performed a validation test.
In this test, we adopted a PHOENIX model spectrum with $T_{\rm eff} = 4700$ K, log$g = 4.5$, and [Fe/H]$= -0.5$ as the template.  
First, we generated a series of synthetic spectra by convolving this template with rotational broadening profiles spanning $v{\rm sin}i = $ 1 km/s to 50 km/s in steps of 1 km/s.
Next, Gaussian noise was added to each synthetic spectrum, drawn using the median of the flux uncertainties from the observed MRS.
Finally, we measured $v{\rm sin}i$ from both the blue and red bands of each synthetic spectrum using the same procedure described in Section \ref{sec:vsini}.
The results show that the recovered $v{\rm sin}i$ values from the blue and red bands are nearly identical and {\bf consistent with} the input values (Figure \ref{vsini_test.fig}).  
This indicates that the $v{\rm sin}i$ measurement ($45.5\pm14.4$ km/s) from the blue band for G6096 is reliable within $1\sigma$ confidence.  
Moreover, the discrepancy in the $v{\rm sin}i$ values between the blue and red bands is likely real.
However, it should be noted that the best-fit $v{\rm sin}i$ does not match the observed spectra well (Figure \ref{best_fit_vsini.fig}), suggesting the possible presence of an additional visible component.

\section{RV and astrometric model}  
\label{sec:RV_Astrometry}

\setcounter{figure}{0}
\setcounter{table}{0}

For the RV model, the variation of stellar RVs due to a companion at epoch $t_j$ is
\begin{equation}
  \hat{v}_{j}= K[\cos(\omega+\nu(t))+e\cos(\omega)]~,
\end{equation}
where $\nu(t)$ is the true anomaly and is related to the eccentric anomaly, $E(t)$, which is given by 
\begin{equation}
    {\rm tan}\frac{\nu(t)}{2}=\sqrt{\frac{1+e}{1-e}}\cdot {\rm tan}\frac{E(t)}{2}~.
\end{equation}
This relation can be derived geometrically. $E(t)$ can be solved using Kepler’s equation
\begin{equation}
    \frac{2\pi}{P}(t-T_{p})=E(t)-e\,{\rm sin}\,E(t)~,
\end{equation}
where $T_p$ is the time of periastron passage.

Therefore, the likelihood for the measured RV ($v_{j,k}$) using different instruments can be calculated by
\begin{equation}
  \mathcal{L}_{\rm RV} = \prod_{j=1}^{N_{\rm RV}} \prod_{k=1}^{N_{\rm inst}}
    \frac{1}{\sqrt{2\pi(\sigma_{j,k}^2+\sigma_{{\rm jit},k}^2)}}
    \exp\!\left(-\frac{(v_{j,k}-\hat{v}_{j,k}-\gamma_k)^2}
    {2(\sigma_{j,k}^2+\sigma_{{\rm jit},k}^2)}\right)~.
\end{equation}
where $N_{\rm RV}$ and $N_{\rm inst}$ are respectively the number of RV measurements and instruments, and $\gamma_k$ and $\sigma_{{\rm jit},k}$ are respectively the RV offset and the so-called ``RV Jitter'' for different instruments. 

Gaia NSS orbital solution is typically expressed using four TI coefficients and three Campbell elements. We treated them, along with the parallax, as observations and construct them as an eight-parameter vector, $\vec{\iota}$, 
\begin{equation}
\{A,\,B,\,F,\,G,\,P,\,e,\,t_p,\,\varpi\}.
\end{equation}
The TI coefficients are defined in angular units, i.e.,
\begin{align}
    A &= a_0\,(\cos\omega\,\cos\Omega - \sin\omega\,\sin\Omega\,\cos i), \\
    B &= a_0\,(\cos\omega\,\sin\Omega + \sin\omega\,\cos\Omega\,\cos i), \\
    F &= -a_0\,(\sin\omega\,\cos\Omega + \cos\omega\,\sin\Omega\,\cos i), \\
    G &= -a_0\,(\sin\omega\,\sin\Omega - \cos\omega\,\cos\Omega\,\cos i),
\end{align}
where $a_0$ is the semi-major axis of the system's photocenter and is related to $a_\star$ by
\begin{equation}
\label{equ:photo}
a_{0} = a_\star\,\varpi\,(1-\frac{m_\star\,f_c}{m_c\,f_\star})(1+\frac{f_c}{f_\star})^{-1}.  
\end{equation}
For dark companion ($f_c=0$), we have $a_0=a_\star\,\varpi$. For luminous companion, we introduced an extra free parameter, $f_{\rm ap}$, to indicate the mass-flux term, i.e., $a_0=a_\star\,\varpi\,f_{\rm ap}$. This simplification is convenient and makes our model independent from any unreliable assumption of component masses.  

Finally, the likelihood for a Gaia two-body solution can be written as
\begin{equation}
  \mathcal{L}_{\rm gaia}=
  \frac{1}{\sqrt{(2\pi)^8|\Sigma|}}
  {\rm exp}\left(-\frac{1}{2}(\hat{\vec{\iota}}-\vec{\iota}\,)^{T}[\Sigma]^{-1}(\hat{\vec{\iota}}-\vec{\iota}\,)    \right)~,
\end{equation}
where $\vec{\iota}$ and $\hat{\vec{\iota}}$ are, respectively, the eight-parameter vector for Gaia measured and model-calculated values, and $\Sigma$ is the associated covariance compiled from the \texttt{nsstools} code\footnote{\url{https://gitlab.obspm.fr/gaia/nsstools}} \citep{Halbwachs2022}.  
Table \ref{Tab:result} lists the best-fit results derived from the joint RV and astrometric analysis.



\begin{table*}[!]
\centering
\caption{The best-fit results derived from {\it The Joker} and the joint fitting of RV and astrometry.}\label{Tab:result}
\begin{tabular*}{\textwidth}{@{}@{\extracolsep{\fill}}llccccc@{}}
\hline \hline
Parameter & Description & RV Only & Gaia Only & \multicolumn{2}{c}{RV+Gaia} & Prior${\rm ^a}$ \\
&&{\it The Joker}&&Dark&Luminous\\
\hline
$P$&Orbital period (days)&${447.18}_{-3.16}^{+3.35}$&${451.9}_{-2.5}^{+2.5}$&${443.52}_{-1.0}^{+0.86}$&${444.42}_{-0.77}^{+0.79}$&Log-$\mathcal{U}(-1,16)$\\
$K$&RV semi-amplitude (km s$^{-1}$)&${17.74}_{-0.14}^{+0.16}$&---&${17.20}_{-0.38}^{+0.37}$&${17.47}_{-0.39}^{+0.37}$&Log-$\mathcal{U}(-1,16)$\\
$e$&Eccentricity &${0.18}_{-0.02}^{+0.03}$&${0.194}_{-0.063}^{+0.063}$&${0.134}_{-0.019}^{+0.020}$&${0.107}_{-0.018}^{+0.019}$&$\mathcal{U}(0,1)$\\
$\omega$&Argument of periapsis ($^\circ$)${\rm ^b}$&$342.6^{+9.2}_{-13.8}$&${240.2}_{-8.4}^{+8.4}$&${282.9}_{-1.8}^{+1.9}$&${262.0}_{-8.3}^{+7.8}$&$\mathcal{U}(0,360)$\\
$M_{0}$&Mean anomaly ($^\circ$)&$41.8^{+7.4}_{-8.0}$&---&${224.4}_{-4.9}^{+3.9}$&${245.9}_{-7.6}^{+7.5}$&$\mathcal{U}(0,360)$\\
$\Omega$&Longitude of ascending node ($^\circ$)&---&${233.5}_{-1.5}^{+1.5}$&${225.7}_{-2.3}^{+2.4}$&${237.1}_{-4.0}^{+4.3}$&$\mathcal{U}(0,360)$\\
$i$&Inclination ($^\circ$)&---&${64.4}_{-2.2}^{+2.2}$&${70.84}_{-0.90}^{+0.87}$&${64.0}_{-1.4}^{+1.4}$&Cos$i$-$\mathcal{U}(-1,1)$\\
$\varpi$&Parallax (mas)&---&${3.244}_{-0.055}^{+0.055}$&${3.182}_{-0.050}^{+0.049}$&${3.201}_{-0.043}^{+0.043}$&$\mathcal{U}(10^{-6},10^{6})$\\
$f_{\rm ap}$&Mass-flux term&---&---&---&${0.668}_{-0.055}^{+0.072}$&$\mathcal{U}(10^{-6},10^{6})$\\
\hline
$T_{\rm p}$ &Periapsis epoch (JD-2457000)&---&${491}_{-12}^{+12}$&${112.0}_{-5.2}^{+6.5}$&${84.9}_{-9.4}^{+10}$&---\\
$a_{0}$ &Photocenter amplitude (mas)&---&${1.624}_{-0.041}^{+0.041}$&${2.338}_{-0.057}^{+0.060}$&${1.69}_{-0.10}^{+0.14}$&--- \\
$A$&Thiele–Innes coefficient (mas)&---&${-0.01}_{-0.16}^{+0.16}$&${-0.897}_{-0.077}^{+0.067}$&${-0.50}_{-0.14}^{+0.14}$&---\\
$B$&Thiele–Innes coefficient (mas)&---&${1.01}_{-0.14}^{+0.14}$&${0.147}_{-0.072}^{+0.068}$&${0.60}_{-0.14}^{+0.14}$&---\\
$F$&Thiele–Innes coefficient (mas)&---&${-1.118}_{-0.027}^{+0.027}$&${-1.467}_{-0.084}^{+0.086}$&${-0.99}_{-0.11}^{+0.10}$&---\\
$G$&Thiele–Innes coefficient (mas)&---&${-0.93}_{-0.14}^{+0.14}$&${-1.747}_{-0.080}^{+0.075}$&${-1.35}_{-0.15}^{+0.14}$&---\\
\hline
$\gamma_{\rm LRS}$&RV offset for LRS (km\,s$^{-1}$)&&---&${5.66}_{-0.39}^{+0.41}$&${6.19}_{-0.42}^{+0.46}$&$\mathcal{U}(-10^{6},10^{6})$\\
$\gamma_{\rm MRS}$&RV offset for MRS (km\,s$^{-1}$)&&---&${5.23}_{-0.23}^{+0.25}$&${5.86}_{-0.33}^{+0.36}$&$\mathcal{U}(-10^{6},10^{6})$\\
$J_{\rm LRS}$&RV jitter for LRS (km\,s$^{-1}$)&&---&${0.69}_{-0.48}^{+0.85}$&${0.50}_{-0.35}^{+0.59}$&$\mathcal{U}(-10^{6},10^{6})$\\
$J_{\rm MRS}$&RV jitter for MRS (km\,s$^{-1}$)&&---&${1.11}_{-0.20}^{+0.25}$&${1.28}_{-0.19}^{+0.25}$&$\mathcal{U}(-10^{6},10^{6})$\\
\hline
BIC&Bayesian information criterion&&&660&632&\\
\hline
\multicolumn{7}{l}{$\rm ^a$ Log-$\mathcal{U}(a, b)$ is the logarithmic uniform distribution between $a$ and $b$, Cos$i$-$\mathcal{U}(a, b)$ is the cosine uniform distribution of $i$}\\
\multicolumn{7}{l}{$\rm ^b$ The argument of periastron of the stellar reflex motion, differing by $\pi$ with companion orbit, i.e., $\omega_{\rm c}=\omega+\pi$.}.\\
\end{tabular*}
\end{table*}

\section{Joint fitting procedure for SED and $f_{\rm ap}$}  
\label{sec:joint_fit_sed_fap}

\setcounter{figure}{0}
\setcounter{table}{0}

First, we constructed a grid based on stellar evolutionary tracks derived from PARSEC models\footnote{\url{https://gitlab.obspm.fr/gaia/nsstools}}.
This grid covers age, mass, radius, effective temperature, surface gravity, metallicity, and predicted magnitudes in the {\it Gaia} ($G$, $BP$, and $RP$), the Two Micron All Sky Survey ({\it 2MASS}; $J$, $H$, and $K_{\rm S}$), and the Wide-field Infrared Survey Explorer ({\it WISE}; $W$1 and $W$2).
Using this grid, we generated a set of hierarchical triple systems to perform the joint fit.

In the joint fitting procedure, we assumed that the primary ($M_1$), secondary ($M_2$), and tertiary ($M_3$) share a common age and metallicity.
The mass function and inclination were fixed to $f(M_2) = 0.246 M_{\odot}$ and $64^{\circ}$, respectively.
Only the age, metallicity, $M_1$, and $M_2$ were treated as free parameters. 
The value of $M_3$ was then derived from the mass function, orbital inclination, and the values of $M_1$ and $M_2$.
Moreover, we adopted uniform priors for the logAge, metallicity, $M_1$ and $M_2$, with ranges of [9, 10.1] and [-1, 1], [0.1, 1] $M_{\odot}$, and [0.1, 1] $M_{\odot}$.
An MCMC sampler with 10 walkers and 10,000 iterations was employed to perform the joint fitting.
The resulting best-fit parameters are summarized in Table \ref{joint_fitting.tab}.


\section{Comparison of inner and outer periods in known triple systems}  
\label{sec:Pin_Pout}

\setcounter{figure}{0}
\setcounter{table}{0}

\begin{figure}
    \center
    \includegraphics[width=0.49\textwidth]{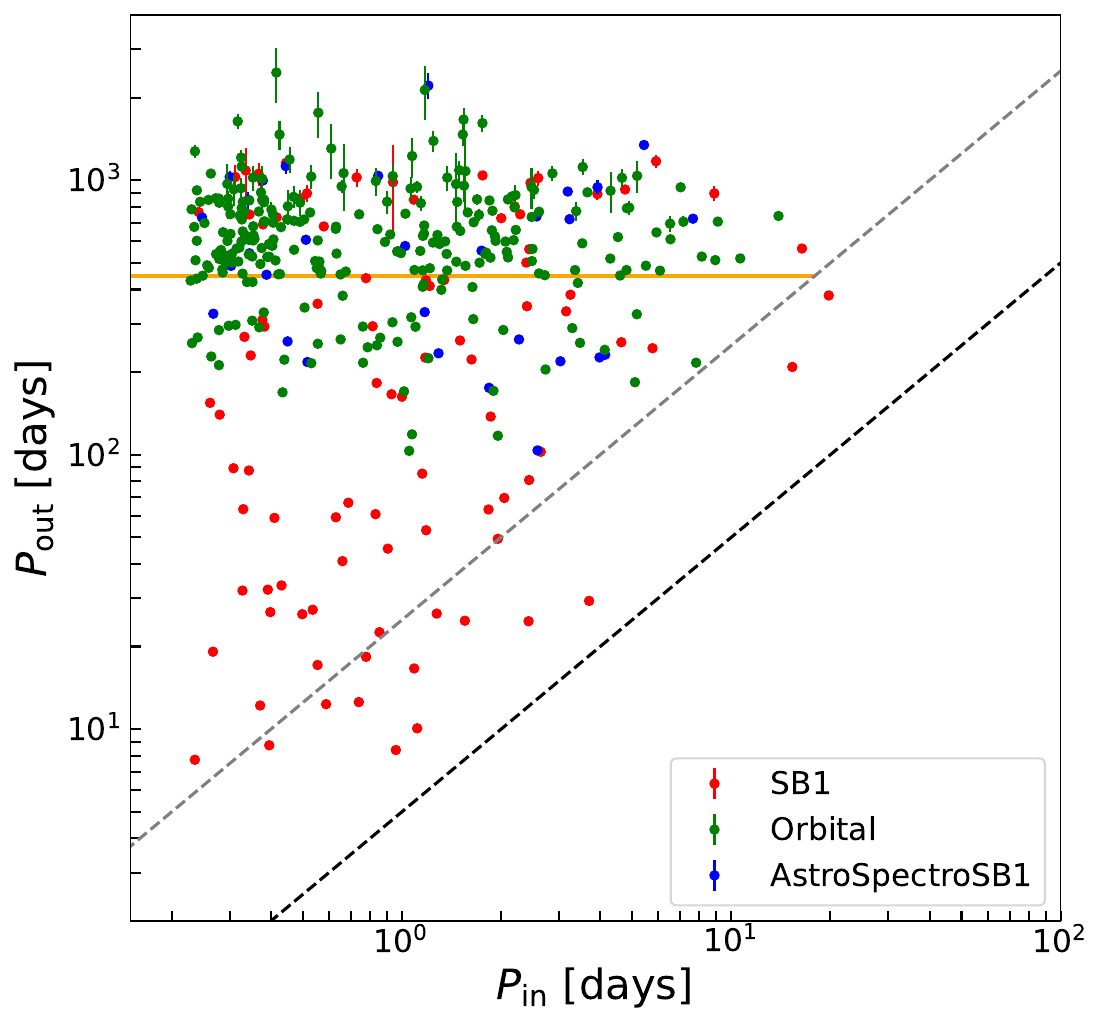}
    \caption{Comparison of inner ($P_{\rm in}$) and outer ($P_{\rm out}$) orbital periods for triple systems from \cite{2023A&A...670A..75C}. The orange shaded region represents the orbital period range of G6096 derived by {\it The Joker}. The black and gray dashed lines denote the relations $P_{\rm out}/P_{\rm in} = 5$ and $P_{\rm out}/P_{\rm in} = 25$, respectively.}   
    \label{compare_3b_systems.fig}
\end{figure}

Figure \ref{compare_3b_systems.fig} shows a comparison of the inner orbital period ($P_{\rm in}$) and the outer orbital period ($P_{\rm out}$) for triple systems from \cite{2023A&A...670A..75C}.

\end{document}